\documentclass[conference]{IEEEtran}

\usepackage[utf8]{inputenc} 
\usepackage[T1]{fontenc}    
\usepackage{url}            
\usepackage{amsfonts}       
\usepackage{nicefrac}       
\usepackage{microtype}      
\usepackage{xcolor}         
\usepackage{amstext}        
\usepackage{tikz}           
\usepackage{subcaption}     
\usepackage{amsmath}
\usepackage{adjustbox}

\usetikzlibrary{positioning, arrows}

\newcommand{\Z}{{\mathbb Z}}
\newcommand\independent{\protect\mathpalette{\protect\independenT}{\perp}}
\def\independenT#1#2{\mathrel{\rlap{$#1#2$}\mkern2mu{#1#2}}}

\usepackage{needspace}
\usepackage[absolute,overlay]{textpos} 

\begin{document}

\title{Graphical Causal Reasoning for Root Cause Analysis in Cloud Networks}

\author{%
Fabien Chraim \\
\texttt{chraim@amazon.com} \\
Amazon Web Services\\
\and
Dominik Janzing \\
\texttt{janzind@amazon.de} \\
Amazon Web Services\\
\and
John Evans \\
\texttt{jevanamz@amazon.co.uk} \\
Amazon Web Services
}

\maketitle

\begin{textblock*}{0.9\textwidth}(0.12\textwidth,1.1\textheight)
\centering
\footnotesize
© 2026 IEEE. This is the author's preprint version of a paper accepted for publication in the Proceedings of IEEE International Conference on Communications, 2026. The final version will appear in IEEE Xplore.
\end{textblock*}

\begin{abstract}
Cloud-computing relies on large-scale networks which are inherently complex systems. In this paper, we present a novel approach to root cause analysis (RCA) of cloud network incidents, leveraging graph-based causal discovery techniques. Our method addresses the limitations of rule-based automation by introducing a spatiotemporal grouping strategy and an automation ontology to reduce the dimensionality of the problem. We construct a causal graph from binary time series data using bivariate Granger causality and conditional independence tests. For inference, we introduce a probabilistic method that assigns edge-specific conditional probabilities as a function of time lag, allowing for interpretable, time-aware root cause scoring via causal graph traversal.

We evaluated the system using a labeled dataset of 35 production incidents from a major cloud provider. The model successfully recalled the correct root cause in 85.7\% of incidents and produced an exact match in 74.3\%. In production, the deployed system has been used in over 800 real-world incidents, with positive qualitative feedback from network engineers. These results highlight the practicality of a data-driven, causal approach to RCA in dynamic and large-scale operational environments.

\end{abstract}

\section{Introduction}
Modern cloud computing environments are underpinned by large-scale networks, designed with layers of redundancy to withstand failures. Multiple components might degrade or fail without immediately compromising the overall system. Yet, even in these resilient systems, failures are inevitable. When they occur, they generate a variety of causal and symptomatic signals, often simultaneously. This makes root cause analysis (defined as the process of determining which cause to address first to restore a service) an operational challenge. There is often not a single root cause, but rather a set of interrelated issues \cite{cook1998complex}.

Modern hyperscale networks generate vast volumes of operational signals, making management increasingly complex and costly. Rule-based automation struggles to scale, often failing to capture dynamic, topology-aware interactions. As a result, operators must constantly maintain brittle rules and escalate unresolved issues to specialists, leading to delays and increased operational burden.

To address these challenges, we propose a graphical causal framework for discovery and inference in complex network environments. This framework goes beyond rule-based automation by uncovering underlying causal relationships between faults, risks, and actions, enabling more accurate and scalable RCA. Central to our approach are a spatiotemporal grouping strategy and a network automation ontology. The grouping method narrows the causal discovery search space by limiting variable consideration to those that co-occur within spatial and temporal proximity during network incidents. The automation ontology categorizes network layers, components, faults, and actions, providing structure to the data and enabling the model to scale across hyperscale environments.

Using six months of incident data from a major cloud provider, we construct a causal graph that captures statistical dependencies and directional interactions between variables.
For inference, we develop a probabilistic graph traversal method that scores candidate root causes by aggregating edge-specific conditional probabilities as a function of time lag. This allows the system to reason about causal propagation over time and select root causes with the highest temporal support.

We deployed this framework as an API integrated into the cloud provider's operational tooling. Over a seven-month period, the system was used in more than 800 real-world incidents. To evaluate its accuracy, we conducted a blind assessment on 35 labeled incidents and found that our model recalled the correct root cause in 85.7\% of cases (Recall@3), and produced an exact match in 74.3\%. These results demonstrate both the practicality and effectiveness of combining causal discovery with probabilistic, time-aware inference for root cause analysis (RCA) at scale.

\section{Background and Related Work}

Traditional approaches to RCA in complex systems often rely on comparing normal and outlier regimes, where a distribution shift or anomaly triggers root cause identification based on deviation from expected behavior. In this setting, RCA techniques typically analyze differences in metrics or event distributions to locate the source of failure. Some recent systems structure telemetry data into graphs for interactive RCA and are based on statistical similarity and require rich time-series features \cite{hardt2023petshop}.

Causal inference methods like RCD (root cause discovery) \cite{ikram2022root} extend the PC (Peter-Clark) algorithm to detect nodes with shifted conditional distributions, while CIRCA (causal inference-based RCA) \cite{li2022circa} tests for distribution changes across known causal structures. Recent theoretical work such as Li et al. \cite{li2025rootcausediscoverypermutations} addresses root cause discovery under linear structural equation models using a single interventional sample against a background of observational data. Their method (which assumes linearly related continuous data) uses permutation-based Cholesky decomposition to identify the intervened variable in a high-dimensional setting and is motivated by gene expression analysis. However, these methods assume clear pre/post-failure segmentation and lack time-aware modeling. Microservice-focused approaches like CausalRCA \cite{xin2023causalrca} and BARO \cite{pham2024baro} use deep generative models and Bayesian change point detection respectively, but face scalability challenges and don't model lagged effects. DCD-FG (differentiable causal discovery of factor graphs) \cite{lopez2022dcdfg} targets scale using differentiable optimization over low-rank DAGs, but assumes global low-rank structure unsuitable for heterogeneous network environments. Counterfactual approaches \cite{root_cause_analysis,okati2024} require well-characterized causal models difficult to obtain in dynamic infrastructure.

In contrast, our method assumes only binary anomaly indicators over time (e.g., fault present or not), and operates entirely within the anomalous regime, without needing a baseline of "normal" data. We learn a causal graph over these binary indicators and perform inference directly on this structure. To our knowledge, this scenario which uses long discretized binary histories of outlier signals without relying on normal-vs-anomaly comparison, is not well represented in the RCA literature.

The closest work \cite{aumayr2019probabilistic} uses probabilistic reasoning over knowledge graphs from troubleshooting documents, but lacks temporal propagation modeling and direct causal structure discovery. Our work differs by integrating graph-based causal discovery with spatiotemporal grouping and network automation ontology for hyperscale environments. We construct causal graphs directly from binary time series using Granger causality and conditional independence tests, introducing probabilistic inference with edge-specific conditional probabilities as functions of time lag for interpretable, scalable, time-aware reasoning.

\section{Model-Based Automation Approach}\label{sec:causal-discovery}
Our approach to automated network operations relies on a causal graph which captures the cause-effect relationship between our variables. To extract this graph, we first apply Granger causality \cite{granger1969investigating} between variable pairs, then further refine the graph with conditional independence tests applied to variable triplets. In this section we explore the mathematical framework behind our approach.

\subsection{Bivariate Granger Causality} \label{sec:bivariate}
For a pair of variables $(A, B)$ we are interested in answering the question: does $B$ cause $A$?
Since our observations come as timeseries which show correlations across timestamps, the core of our approach
will be Granger causality \cite{granger1969investigating}, the standard method for causal timeseries analysis.  
If the observations from $A:=(A_t)_{t\in \Z}$ and $B:=(B_t)_{t\in \Z}$ are not subjected to hidden common causes, $B$ has a causal influence on $A$ whenever the past of $B$ helps in predicting $A$ from its own past (see chapter 10 of \cite{causality_book} for conditions that guarantee that Granger causality finds all causal relations). Denoting 
the past observations as $B_{\rm past(t)}:= B_{t-1}, B_{t-2},...$, and conditional statistical independences by $\independent$, we say that $B$ Granger causes $A$ whenever
$
A_t \not\independent B_{\rm past(t)} \,| A_{\rm past(t)}.   
$

In section \ref{sec:ontology} below, we will describe our data in more detail. For now, we mention that it is \textit{event}-based since we derive it from existing automation systems running on the network (alerts, workflow steps, etc.). This data is therefore binary in nature: at a time $t$, a variable $A$ takes a value of $1$ if its underlying event occurred, and a value of $0$ otherwise. We are therefore looking for predictors on our binary timeseries. We employ logistic regression for this task. To test whether $B$ Granger causes $A$, we define a restricted model using the $A$ variable alone, and a second, unrestricted model using both $A$ and $B$

\begin{align} \label{eq:restrictedunrestricted}
\log\left(\frac{P(A_t = 1)}{1-P(A_t=1)}\right) &= \alpha_0 + \alpha_1 A_{past(t)} \tag{Restricted Model} \\ 
\log\left(\frac{P(A_t = 1)}{1-P(A_t=1)}\right) &= \beta_0 + \beta_1 A_{past(t)} + \beta_2 B_{past(t)} \tag{Unrestricted Model}
\end{align}

Once the restricted and unrestricted models are fitted, we are ready to test whether the addition of $B$ significantly improves the predictive power over $A$. From the models, we extract the log likelihoods ($LL_{restricted}$ and $LL_{unrestricted}$) and the degrees of freedom ($DoF_{restricted}$ and $DoF_{unrestricted}$). We define the likelihood ratio statistic as $\lambda = -2 \times (LL_{restricted} - LL_{unrestricted})$. We also take the difference in degrees of freedom to prepare for a significance test under the $\chi^2$ distribution: $k = DoF_{unrestricted} - DoF_{restricted}$.

To test whether the goodness of fit difference is significant under that distribution, we derive the $p\text{-value}$ for the test as follows.
\begin{equation} \label{eq:pvalue-chisq}
p\text{-value} = 1-F_{\chi^2_k}(\lambda)
\end{equation}

This test is designed to ensure that, even though the unrestricted model may have more degrees of freedom, its goodness of fit has to be \textit{significantly} better than the restricted model. Significance in this case can be determined by setting a threshold on the $p\text{-value}$ of equation \ref{eq:pvalue-chisq} (commonly 0.05 \cite{di2020statistical}). Practically speaking, if we find that our $p\text{-value}$ is below this threshold, we can add a causal edge from $B$ to $A$, signifying that $B$ may Granger cause $A$. We repeat this procedure for all pairs of variables and obtain an initial causal graph.

\subsection{Conditional Independence}\label{sec:triplet}

While the pairwise evaluation provides an initial approximation of the causal graph, it is limited in its ability to capture more complex dependencies within a system. For instance, it can mistake indirect links as direct ones, which misguides the reconstruction of the causal pathway. To better understand the causal pathways and refine the graph, a more comprehensive triplet analysis is necessary. This involves assessing conditional independence, which helps to uncover indirect effects and mediation relationships that are otherwise hidden in pairwise comparisons, leading to a more nuanced and accurate representation of the underlying causal structure. 

To this end, we propose the following steps. After having identified an arbitrary ancestor $B$ of our target variable $A$ (i.e., a variable influencing $A$ directly or indirectly), we search for a third variable $C$ which is an ancestor of $B$. To tell whether the influence of $C$ on $A$ is direct or mediated by $B$, we test $C_{\rm past(t)} \independent A_t\,| B_{\rm past(t)}$. If this is the case, we assume that the three variables $C,B,A$ are linked by the causal DAG $C\to B \to A$, otherwise, both $B$ and $C$ are parents of $A$. In these considerations we have implicitly assumed that the triplet $A,B,C$ is causally sufficient, which collides slightly with our knowledge that they are heavily linked to the remaining variables. We also excluded the case where one of the variables in the triple is a common cause of the others. Further, we are assuming that the causal ``summary graph'' \cite{NIPS2013_47d1e990} (i.e. the graph modelling each time series as a node) is acyclic on $A,B,C$. We also exclude instantaneous influence. Otherwise the above conditional independence tests cannot necessarily distinguish between direct and indirect effects. Then we can have $A_t \not\independent C_{\rm past(t)}\,| B_{\rm past(t)}$ even when $B$ influences $A$ via $B$, as one can easily check using d-separation in causal Bayesian networks \cite{Pearl:00}. All three assumptions are certainly oversimplifications, but without them the causal analysis would not scale to our large environment. We get some support for our assumptions by observing multiple conditional independences in the data (section \ref{sec:discovery}), which could not be the case if the violations would be too severe.

\section{Causal Discovery At Scale} \label{sec:discovery}
Applying the machine learning framework from Section \ref{sec:causal-discovery} to cloud networks is challenging due to the sheer scale and complexity: millions of devices and thousands of signal types result in an unmanageably large variable space. Exhaustive causal reasoning over all combinations is computationally infeasible. To address this, we introduce an automation ontology to simplify the space, and a spatiotemporal grouping strategy to control combinatorial growth.

\subsection{Automation Ontology for Network Operations}\label{sec:ontology}

The automation ontology presented here is designed to simplify and enhance the management of both large- and small-scale computer networks. Its modular structure and focus on capturing causal relationships between observations, faults, risks, and actions allow for generalization to a wide variety of complex systems (e.g., service-oriented architectures).

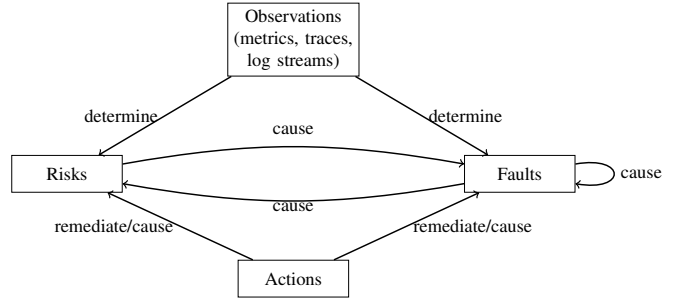
\begin{figure}[!htbp]
\centering
\resizebox{\linewidth}{!}{
\begin{tikzpicture}[
  box/.style={rectangle, draw, minimum height=2em, minimum width=6em, text centered, align=center},
  arrow/.style={->, thick}
]

\node[box] (obs) {Observations \\ (metrics, traces, \\ log streams)}; 
\node[box, below left=1.5cm and 2cm of obs] (risks) {Risks};          
\node[box, below right=1.5cm and 2cm of obs] (faults) {Faults};        
\node[box, below=3.5cm of obs] (actions) {Actions};                     

\draw[arrow] (obs) -- (risks) node[midway, left] {determine};
\draw[arrow] (obs) -- (faults) node[midway, right] {determine};

\draw[arrow] (risks) to[out=10, in=170] (faults) node[midway, above, yshift=-55] {cause}; 
\draw[arrow] (faults) to[out=-170, in=-10] (risks) node[midway, below, yshift=-85] {cause};  
\draw[arrow, loop right, out=10, in=-10, distance=1cm] (faults) to node[right] {cause} (faults);


\draw[arrow] (actions) -- (faults) node[midway, right] {remediate/cause};
\draw[arrow] (actions) -- (risks) node[midway, left] {remediate/cause};

\end{tikzpicture}
}
\caption{Automation Ontology: Observations (metrics, traces, log streams) determine risks and faults. Risks can materialize into faults, and faults can generate risks. Actions can remediate or cause both faults and risks.}
\label{fig:ontology}
\end{figure}

Figure \ref{fig:ontology} shows a graphical representation of the elements of the ontology as nodes, and their relationships as edges. In many computer systems, observations consist of metrics, traces, and log streams, which can be used to detect faults and assess risks via threshold-based or anomaly detection mechanisms. Actions, such as deployments and configuration changes, enable operators to mitigate or remediate these faults and risks. However, these same actions can inadvertently introduce new faults or risks when they contain bugs or trigger unexpected conditions. The self-loop on the `Faults' node illustrates the cascading nature of faults within complex systems.

To adapt this ontology for network automation, we defined 14 fault categories (e.g., hardware errors, control plane issues, packet loss), 11 action categories (e.g., device replacement, traffic engineering), and grouped over 800 alerts and 10,000 workflow steps accordingly.

Topological information plays a key role: faults propagate across multiple hops in both physical and logical topologies. Cloud networks replicate layered architectures (e.g., 3-tier Clos \cite{clos1953study,hennessy2017computer}) across geographies, grouping devices by role. We model this structure using 185 layers, treating the layer as a dimension in the variable space. This yields $185 \times (14 + 1 + 11) = 4,810$ binary variables, where each (layer, variable) pair is distinct and indicates the presence or absence of some fault, risk or action. Although the layer-based abstraction reduces complexity, brute-force causal discovery over $4{,}810^2$ variable pairs remains intractable. The necessary pruning approach is described next.

\subsection{Incident Creation with Spatiotemporal Grouping}\label{sec:rumorspread}
In large-scale cloud networks, where automation fails to detect or remediate an issue, \textbf{escalated} events are generated for manual intervention. These escalations require operators to diagnose and resolve the issue, often without a full understanding of the context that led to the event.

To reconstruct incident context, we group signals spatiotemporally. Starting from the affected locations, we collect faults, risks, and actions within $X$ hops and a $Y$-minute window (these are configurable parameters spanning local topology and time). This defines an incident: the unit we aim to diagnose. Some collected signals may be unrelated, and $X$, $Y$ must balance capturing relevant context while minimizing noise. We exclude raw metrics, traces, and logs, which are summarized in the signal data and are difficult to scale.

Not only does this method allow us to gather the full context needed for accurate diagnosis and RCA, but it also helps reduce the size of the variable space in two ways: a) we merge signals which map to the same layer and ontological category together, and b) we only need to test combinations of variables that occur within the same incident, rather than across the entire network.

\subsection{Causal Graph}\label{sec:cgvalidation}
We collected a dataset of escalated network events from January to June 2024. After spatiotemporal grouping and ontology mapping, we extracted 25,474 incidents yielding 76,595 variable pairs for bivariate Granger analysis (Section \ref{sec:bivariate}). This substantial reduction in the number of possible combinations is achieved by eliminating pairs which did not occur in any incident. This produced 1,681 causal edges (2.19\%), which expanded to 1,989 edges after conditional independence tests (Section \ref{sec:triplet}). We allowed bidirectional edges to account for timing inaccuracies and missing signals which may be the true root cause.

\begin{figure}[h!]
\centering
\includegraphics[width=0.265\textwidth]{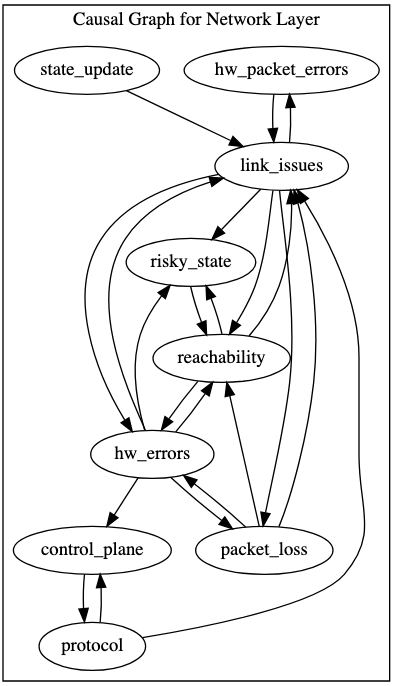}
\caption{Subset of the learned causal graph for a particular network layer.}
\label{fig:cg-subset}
\end{figure}

In figure \ref{fig:cg-subset}, we show a subset of our causal graph that is relevant to a single network layer (excluding edges to and from adjacent layers). Many of the causal relationships discovered align with domain knowledge. For instance, hardware errors (hw\_errors) on network routers do in fact cause both dataplane packet loss (packet\_loss) and control plane issues. Additionally, state updates can lead to link issues such as flapping or downtime. However, some discovered edges contradict domain knowledge. For example, link issues are shown as causes of hardware errors, which is incorrect: they are typically consequences. This misdirection may result from maturity timers in our detectors (meaning that signals are intentionally delayed to account for transient effects/noise in the metrics). Furthermore, the graph is incomplete; for example, hardware packet errors (e.g., ingress errors on a link) are known to cause dataplane packet loss, but this causal relationship is missing, likely due to insufficient data. Another key observation is that the graph contains cycles, meaning it is not a directed acyclic graph (DAG). The question remains: can this graph still be used for effective RCA?

\section{Probabilistic Root Cause Inference via Causal Path Likelihood} \label{sec:probabilistic-rca}

In this section, we introduce a principled probabilistic inference method for RCA based on traversal of the learned causal graph. The approach leverages time-lagged conditional probabilities and a likelihood scoring mechanism to evaluate candidate root causes.

\subsection{Extracting Time-Lagged Conditional Probabilities}

Once the causal structure of the graph has been learned via pairwise and triplet analysis (Section~\ref{sec:causal-discovery}), we compute a time-dependent conditional probability function for each edge in the graph. Specifically, for an edge $B \rightarrow A$, we define:
\begin{equation}
P_{B \rightarrow A}(\Delta t) = P(A = 1 \mid B = 1, \Delta t = t_A - t_B)
\end{equation}
This function is estimated empirically by computing the frequency of co-occurrence of events $B$ and $A$ at different time lags $\Delta t$ using the training dataset. We discretize time into one-minute bins and estimate $P_{B \rightarrow A}(\Delta t)$ for a horizon of 20 minutes. In Section~\ref{sec:prob-viz}, we illustrate examples of such conditional probability functions.

\subsection{Root Cause Inference by Maximum Path Likelihood}

To determine the most likely root cause for a given incident, we consider each signal in the incident as a candidate root cause $r$. For each candidate $r$, we perform a traversal of the causal graph to identify all directed paths from $r$ to the impact variable $y$ (incidents are generated with the impact variable as the entry point).

Each path $\pi = r \rightarrow v_1 \rightarrow \cdots \rightarrow y$ is assigned a likelihood score based on the product of conditional probabilities along its edges. Let $\Delta t_{uv} = t_v - t_u$ be the time lag between two activated variables $u$ and $v$ on the path. Then the path likelihood is:
\begin{equation}
\mathcal{L}(\pi) = \prod_{(u \rightarrow v) \in \pi} P_{u \rightarrow v}(\Delta t_{uv})
\end{equation}

If $v$ is unobserved but $u$ occurred, we conservatively apply a small constant $\epsilon$ to the edge likelihood.

Since a candidate root cause may have multiple paths to the impact variable, we assign it the maximum path likelihood:
\begin{equation}
\mathcal{L}^*(r) = \max_{\pi \in \Pi_r} \mathcal{L}(\pi)
\end{equation}
where $\Pi_r$ is the set of all paths from $r$ to $y$.

We compute $\mathcal{L}^*(r)$ for all root cause candidates $r$, and rank them by decreasing likelihood. The root causing algorithm returns the top-3 candidates whose maximum path likelihoods exceed a threshold $\theta$.

\subsection{Visualizing Conditional Probability Functions}
\label{sec:prob-viz}

\begin{figure}[h!]
    \centering
    \begin{subfigure}[b]{0.47\textwidth}
        \centering
        \includegraphics[width=\textwidth]{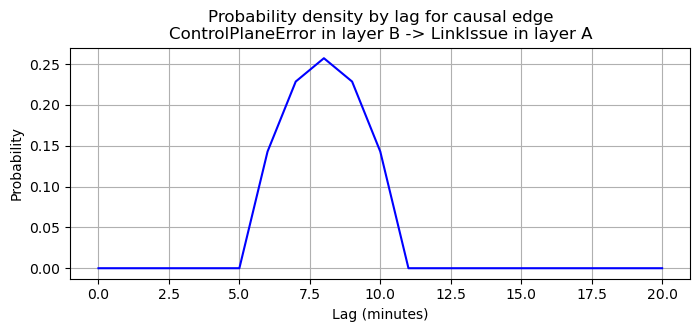}
        \caption{Conditional probability of link issue given control plane error, in two different network layers.}
        \label{fig:controlplanelinkissue}
    \end{subfigure}%

    \begin{subfigure}[b]{0.47\textwidth}
        \centering
        \includegraphics[width=\textwidth]{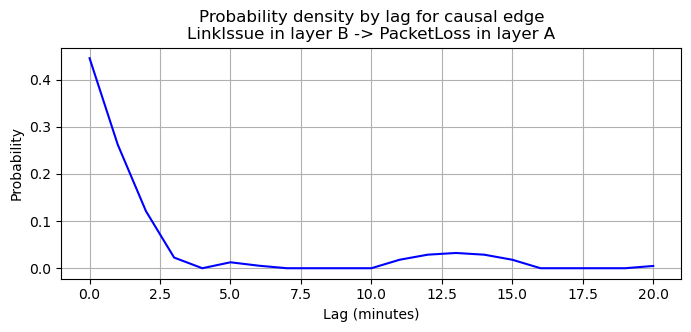}
        \caption{Conditional probability of packet loss given link issue, in two different network layers.}
        \label{fig:linkissuepacketloss}
    \end{subfigure}
    \caption{Example conditional probability functions $P_{B \rightarrow A}(\Delta t)$ extracted from historical incidents. (a) Control plane errors in one layer increasing the likelihood of link issues in another. (b) Link issues increasing the likelihood of packet loss.}
    \label{fig:cond-prob-examples}
\end{figure}

To build intuition about the conditional probability functions $P_{B \rightarrow A}(\Delta t)$, we present two examples in figure~\ref{fig:cond-prob-examples}. Each plot shows the estimated probability of $A$ occurring at a given time lag $\Delta t$ after $B$, over a 20-minute horizon. These examples illustrate how certain types of network events tend to co-occur within specific temporal windows, and how this timing is influenced by the network layer context.

In figure~\ref{fig:controlplanelinkissue}, we observe that control plane errors tend to precede link issues with a characteristic lag of 6 to 10 minutes. Outside this time window, the likelihood of co-occurrence drops significantly, suggesting that such pairs should not be interpreted as causally connected when observed outside the typical lag range. In contrast, figure~\ref{fig:linkissuepacketloss} shows that link issues are most strongly associated with packet loss in the first two minutes following the fault. This indicates a tighter, faster coupling between these event types.

These empirical patterns provide concrete support for the time-lagged conditional probabilities used in our inference method. By capturing both the strength and timing of causal relationships, they enable the system to distinguish meaningful temporal patterns from coincidental co-occurrences, and to reason more effectively about causality during RCA.

\section{Performance in an Operational Environment}

We deployed our model as part of a root cause analysis API used by network engineers in production at a major cloud provider. This API integrates with the internal ticketing system and, upon receiving an incident, performs a spatio-temporal context search, maps signals to ontology concepts, and runs the probabilistic causal inference method described in Section~\ref{sec:probabilistic-rca} to generate root cause hypotheses.

To validate the system's accuracy, we conducted a blind evaluation with network engineers using a labeled dataset of 35 high-impact, low-frequency incidents that bypass existing rule-based automation and escalate to expert operators. Our probabilistic inference method correctly recalled the root cause in 30 out of 35 cases (85.7\%) and provided an exact match in 26 cases (74.3\%). In this case, we use recall@3 which is the proportion of incidents where the ground-truth root cause appears among the top 3 ranked predictions returned by the model.

\begin{figure}[h!]
\centering
\includegraphics[width=0.45\textwidth]{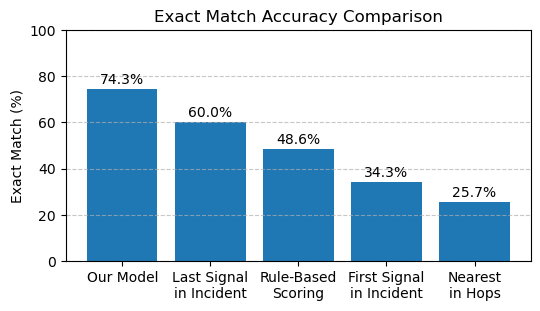}
\caption{A comparison of exact match match accuracy by method.}
\label{fig:exact-match}
\end{figure}

To contextualize these results, we compared them against a rule-based approach designed by expert network engineers, along with three intuitive baselines relying on purely temporal or spatial heuristics (see figure \ref{fig:exact-match}). 

\begin{itemize}
\item The rule-based approach produced 17 exact matches (48.6\%) and correctly recalled the root cause (i.e., with other similarly scored signals) in 22 cases (62.8\%).
\item Selecting the first signal in the incident yielded 12 exact matches (34.3\%).
\item Selecting the last signal (immediately preceding impact) resulted in 21 exact matches (60\%).
\item Selecting the signal nearest to the impact location in terms of network hops produced 9 exact matches (25.7\%) and correctly recalled the root cause (i.e., within the same position) in 18 cases (51.4\%).
\end{itemize}

Unlike the temporal baselines, our method incorporates both temporal and spatial reasoning. And unlike the spatial baseline, which can suffer from ties or local ambiguity, our approach consistently delivers higher precision and recall. These results demonstrate that probabilistic inference over a causal graph provides a significantly more reliable and actionable foundation for automated RCA in large-scale network environments.

Operationally, the model has been used in over 800 real-world incidents over a seven-month period. Engineers provided ratings on a 5-star scale for 48 incidents. Of these, 19 incidents received 5-star ratings (indicating a correct and complete root cause), and 25 received at least 3 stars (partially correct but useful). These results reinforce the practical utility of the system and its ability to guide effective triage in high-pressure environments.

A detailed analysis of the low-rated incidents revealed limitations in our spatial modeling approach. While the spatial simplifications introduced in Section \ref{sec:ontology} successfully removed unnecessary location-specific complexity, they also resulted in excessive information compression. Consequently, the model often equated signals that are inherently different, such as distinguishing faults occurring on the same device from those occurring on sibling or neighboring devices within the same layer. Furthermore, the model struggled with incorporating certain types of signals that could not be scoped to a specific device within a network layer. For example, reachability or route signals scoped to network endpoints and path signals scoped to a sequence of locations on the network remain unsupported.

Addressing these limitations is critical for improving the model's performance. Future work will focus on refining the spatial representation to better capture the nuances of fault and signal relationships, particularly in complex network topologies.

Nevertheless, these evaluation results and the fact that the model was used in hundreds of incidents are a testament to viability of data-driven, causal reasoning methods for automated RCA in complex, real-time operational settings.

\section{Discussion and Future Work}
In this paper, we presented a scalable and interpretable approach to RCA for large-scale network environments, grounded in graph-based causal discovery. By combining spatiotemporal grouping with a network automation-specific ontology, we made causal analysis tractable in settings involving tens of thousands of variables across heterogeneous infrastructure layers. Our discovery process integrates bivariate Granger causality and conditional independence tests to construct a causal graph that captures directional relationships between faults, risks, and actions.

To support practical root cause inference, we introduced a probabilistic method that scores paths through the causal graph using edge-specific conditional probabilities as a function of time lag. This enables the system to reason over observed incident timelines, highlighting candidate root causes based on how well they temporally and structurally explain the impact.

We validated our framework on six months of incident data from a major cloud provider. In a labeled evaluation of 35 real-world incidents, the model recalled the correct root cause in 85.7\% of cases and provided an exact match in 74.3\% (a 25.7\% absolute percentage point improvement over rule-based methods). In production, the system was used in over 800 incidents demonstrating its operational relevance and usability.

While effective, our method has some limitations. In particular, the spatial modeling abstractions (designed for scalability) can lead to information loss when distinguishing between local versus neighboring faults, or when handling metrics scoped to network paths and endpoints. Addressing these challenges is a key direction for future work. Additional opportunities include incorporating explicit temporal constraints into the discovery process and exploring GNN-based models that can learn spatiotemporal structure more flexibly.

Despite these limitations, our approach represents a significant step forward in automated network management, offering a scalable and data-driven alternative to traditional rule-based systems. The framework's ability to discover meaningful causal relationships and provide actionable RCA in a production environment demonstrates its potential to improve network reliability and reduce operational costs in large-scale cloud networks.

\bibliographystyle{unsrt}
\bibliography{stcr_icc}

\end{document}